\documentclass[prl,twocolumn,amsmath,amssymb,floatfix,footinbib,bibnotes]{revtex4-1}

\pdfoutput=1

\usepackage[colorlinks=true,citecolor=blue,linkcolor=magenta]{hyperref}
\usepackage{amsmath}
\usepackage{graphicx}
\usepackage{dsfont}
\usepackage{changepage}
\usepackage{fancyhdr}
\usepackage{amsthm, amssymb}
\usepackage{floatflt}
\usepackage{float}
\usepackage{array}
\usepackage[usenames,dvipsnames]{color}

\newcommand{\be}{\begin{align}}
\newcommand{\ee}{\end{align}}

\def \ua{{\uparrow}}
\def \da{{\downarrow}}
\def \be{\begin{equation}}
\def \ee{\end{equation}}
\def \ba{\begin{array}}
\def \ea{\end{array}}
\def \bea{\begin{eqnarray}}
\def \eea{\end{eqnarray}}
\def \nn{\nonumber}
\def \half{{1\over 2}}

\def \e{{\epsilon}}

\def \a{{\alpha}}
\def \t{{\theta}}

\def \g{{\gamma}}
\def \D{{\Delta}}
\def \d{{\delta}}

\def \s{{\sigma}}
\def \f{{\varphi}}

\def \e{{\epsilon}}

\def \G{{\Gamma}}

\def \av#1{{\langle#1\rangle}}
\def \ket#1{{\,|\,#1\,\rangle\,}}

\def \ba{\begin{align*}}
\def \ea{\end{align*}}

\def \half{{\frac{1}{2}}}

\newcounter{indice}

\def \mrm{\mathrm}

\def \mc{\mathcal}

\begin{document}

\title{Topological States in a One-Dimensional Fermi Gas with Attractive Interactions }

\author{Jonathan Ruhman, Erez Berg and Ehud Altman \\
{\small \em Department of Condensed Matter Physics, Weizmann Institute of Science, Rehovot 76100, Israel}}
\begin{abstract}
{ We describe a novel topological superfluid state, which forms in a one-dimensional Fermi gas with Rashba-like spin-orbit coupling, a Zeeman field and intrinsic attractive interactions. In spite of total number conservation and the presence of gapless excitations, Majorana-like zero modes appear in this system and can be linked with interfaces between two distinct phases that naturally form at different regions of the harmonic trap. As a result, the low lying collective excitations of the system, including the dipole oscillations and the long-wavelength phonons, are doubly degenerate. While backscattering from point impurities can lead to a splitting of the degeneracies that scales algebraically with the system size, the smooth confining potential can only cause an exponentially small splitting. We show that the topological state can be uniquely probed by a pumping effect induced by a slow sweep of the Zeeman field from a high initial value down to zero field. }
\end{abstract}
\maketitle

\emph{Introduction.---} Recent experiments with semiconducting nanowires have shown possible signatures of Majorana zero modes, the hallmarks of a topological superconducting state, localized at the ends of the wires~\cite{Mourik2012,Das2012}. The two key ingredients required to realize such topological states are a single particle dispersion affected by spin-orbit coupling and a Zeeman field, and pairing correlations induced by proximity coupling to an s-wave superconductor \cite{Oreg2010,Lutchyn2010}. 

Systems of ultracold atoms offer a high degree of controllability, and are therefore attractive as platforms for realizing Majorana zero modes~\cite{Jiang2011}. Effective spin-orbit coupling and Zeeman field can also be generated in systems of ultra-cold atoms confined to one dimension\cite{Lin2011,Cheuk2012,Wang2012,Cui2013}.  However, in this case it is much more difficult to induce pairing correlations externally. This naturally leads to the following basic question: can \emph{intrinsic} attractive interactions (generated naturally in atomic systems with Feshbach resonances) lead to a topological phase and Majorana zero modes without externally induced pairing?

If the system was two or three dimensional then attractive interactions, naturally generated in atomic systems with Feshbach resonances, would give rise to a Bardeen-Cooper-Schrieffer (BCS) pairing gap equivalent to that induced by proximity to a bulk superconductor. But this is not the case in the one-dimensional system in question, where long range order superfluid order is impossible. %Moreover, in a system with a single mode at the Fermi energy, even a single-particle gap are established.
Nevertheless, it was shown in
Refs. ~\cite{Fidkowski2011,Sau2011} that proximity coupling of two independent spin orbit coupled wires to a single one-dimensional superconducting wire with quasi-long range pairing correlations would retain a Majorana-like ground state degeneracy. The question remains if a single, isolated wire %with charge conservation 
can sustain similar topological zero modes due to the intrinsic attractive interactions.

\begin{figure*}
\centering
\includegraphics[width=18cm,height=4.5cm]{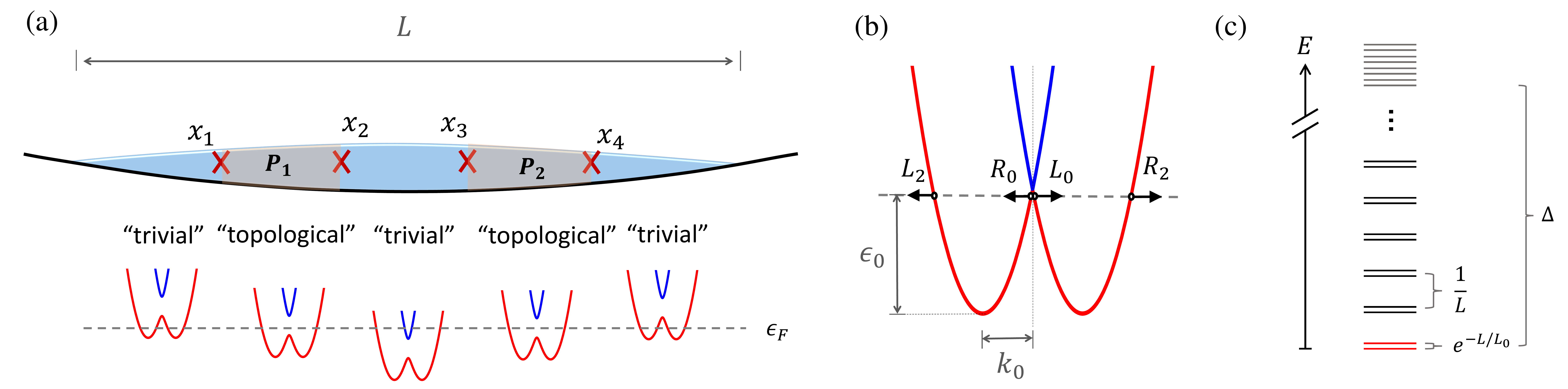}
\caption{(a) A one dimensional Fermi gas with synthetic spin-orbit coupling, a Zeeman field and attractive interactions in a one dimensional harmonic trap.  Majorana zero modes are localized at the interface between topological and trivial regions, approximately where the chemical potential dips below the Zeeman gap at wave vector $q=0$. Two topological segments enumerated by I and II form when the chemical potential at the center of the trap is set to be larger than the Zeeman splitting. When the segments are close to each other there is a finite probability, $\G$, to switch parity between them. (b) The Rashba-like dispersion at $\mu=0$ and $\d_z = 0$ showing our notations of the four modes crossing the Fermi energy. The black arrows denote the spin orientation of the helical modes. (c) Schematic depiction of the energy spectrum showing the topological degeneracies. The low energy excitations associated with dipole oscillations in the trap are spaced by the trap frequency $\Omega\sim 1/L$ as in a conventional system. However, the ground state as well as the collective excitations are doubled (up to the exponentially small splitting) because of a topological degeneracy associated with switching fermion parity between the 'topological' segments. }\label{fig:config}
\end{figure*}

In this paper we use an effective field theory to answer this question and characterize the emergent low energy modes. We show that  this system can exhibit Majorana-like degeneracies in spite of having no proximity coupling to an external pairing field. The zero modes are associated with interfaces between distinct phases that may form in different regions of the trap due to the spatial variation of the chemical potential. We term "topological" the phase established where the chemical potential is inside the Zeeman gap. %, which there dominates over the attractive interactions. 
This phase supports gapless single-fermion excitations. In other regions the attractive interactions dominate and generate a gap to single-fermion excitations. The Majorana-like quasi-zero modes occur in a configuration, as illustrated in Fig. \ref{fig:config}.a, which includes at least two ``topological" regions. The physical picture and observable consequences of the zero modes in this charge conserving system, which emerge from our exact analysis, are notably different from previous mean field studies \cite{Wei2012,Liu2012}. We show how to probe the zero modes and expose their topological origin through a pumping phenomena induced by a quasi-adiabatic sweep of the Zeeman field. 

\noindent{\em Model ---} %As a starting point 
We consider a one dimensional Fermi gas with spin-orbit coupling, a Zeeman field and short ranged attractive interactions described by the following Hamiltonian
\bea
\mc H &=& \int dx\bigg[  \psi^\dag  \bigg( -{\partial_x ^2 \over 2m}+V(x)-\mu +\a\s^x i\partial_x -\d_z \s^z  \bigg) \psi\nn\\
&&~~~~~~~~~- U \, \psi_{\ua}^\dag \psi_{\da}^\dag \psi_{\da}\psi_{\ua}\bigg]\,. \label{H_m}
\eea
Here $\psi_\sigma$ annihilates a fermion with spin $\sigma=\uparrow,\downarrow$, $\psi^T = (\psi_\uparrow,\psi_\downarrow)$, $m$ is the particle mass, $\a$ the spin-orbit coupling strength, $\mu$ the chemical potential, $\d_z$ is an effective Zeeman field, $V(x) = {m \Omega^2} x^2/2$ is the parabolic trapping potential, and $U>0$ is the interaction strength.

The parabolic trap potential can be thought of as a % within this model by a 
position dependent chemical potential. We consider filling the system to a point that the chemical potential in the middle of the trap is located above the Zeeman gap and continuously decreases towards the flanks.
In the usual case where there is a small proximity induced s-wave pairing field $\D<\d_z$, the spatially dependent chemical potential tunes the system from a trivial state in the middle of the trap to two topological states on the sides which further transform to trivial states at the ends of the system (see Fig.~\ref{fig:config}.a).
 An alternative way to tune the system between the same two phases,  is by varying the ratio of $\D/\d_z$, while keeping the chemical potential fixed in the middle of the gap. The topological phase is established in the region where $\D/\d_z<1$.
 %This way of tuning would prove to be a convenient theoretical tool in analyzing the problem without an external pairing field.
This way of tuning proves to be a convenient theoretical tool in deriving the universal low energy theory of the system.

\noindent {\em Low energy theory.---}
As a preparatory step consider an infinite homogenous wire described by the %fully charge conserving 
Hamiltonian (\ref{H_m}) with $\mu=0$. It is convenient to formulate the  long wavelength theory starting from the case with $\d_z=0$. Then we have four fermion modes crossing the Fermi energy, $R_{a}$ and $L_a$, as shown in Fig.~\ref{fig:config}.b. $a=0,2$ labels the modes  at $k=0$ and $k=\pm 2k_0\equiv \pm 2m\a$ respectively. Next, we Bosonize the four chiral modes at $k=0,\pm 2 k_0$: $R_a \sim F_a\sqrt{\rho_0\over 2\pi }\, e^{i(\t_a- \phi_a) }$,  $L_a \sim F_a\sqrt{\rho_0\over 2\pi }\, e^{i(\t_a+ \phi_a) }$, where the commutation relations of the Bosonic fields are given by $[\phi_a(x),\theta_b(x')] = i \pi\, \d_{a,b}\Theta(x-x')$, $F_{0,2}$ are Klein factors to set the Fermionic anti-commutation relations between the modes and $\rho_0$ is the average density.

The Hamiltonian (\ref{H_m}) written in terms of the bosonic fields includes, as usual, a quadratic (Luttinger liquid) part due to the kinetic energy and forward scattering channel of the interaction. On the other hand the Zeeman term and the BCS channel of the attractive interactions give rise to the respective cosine terms  $\int dx \left[g_z\, \cos 2\phi_0 +g_i\, \cos 2\left( \t_0-\t_2 \right) \right]$,
with the coefficients $g_i\approx {\rho_0^2 U \over (2\pi)^2}$ and $g_z \approx {\rho_0 \, \d_z \over 2\pi }$ (at weak coupling). Note that the cosine terms affect both of Luttinger liquid modes $0$ and $2$. However, we can simplify the situation %and decouple one of the gapless Luttinger liquid modes from the cosine perturbations 
by means of the following canonical transformation: % to the basis:
\[\begin{matrix} \;\;\;\;\;\;\;\phi_+ = \phi_0 + \phi_2 \\ \t_+ = \t_2\;\;\; \end{matrix}\;\;\;\begin{matrix} \phi_- = \phi_0\\ \;\;\;\;\;\;\;\t_- = \t_0 - \t_2 \end{matrix}\,,\]
In this representation the two modes are decoupled in the low energy limit\footnote{linear coupling terms $\partial_x \t_- \partial_x \t_+$ and $\partial_x \phi_- \partial_x \phi_+$ generated by the canonical transformation are irrelevant in the long wavelength limit because the mode   ($\t_-,\phi_-$) gapped (except at a critical point).}  and the Hamiltonian takes the form $\mc H = \mc H_+ + \mc H_- $, with
\begin{align}
\mc H_+={u_+\over 2\pi}\int dx &\left[ {{K_{+}}(\partial_x \t_+)^2+{1\over K_+}(\partial_x \phi_+)^2 }\right]\label{H+} \\
\mc H_-={u_-\over 2\pi}\int dx &\left[ {{K_{-}}(\partial_x \t_-)^2+{1\over K_-}(\partial_x \phi_-)^2 }\right] \label{H-}\\
-&\int dx \left[ g_z\, \cos 2\phi_- + g_i \, \cos 2\t_- \right],\nn
\end{align}
and where $u_\pm$ and $K_\pm$ are the renormalized velocities and Luttinger parameters in the two channels. The Hamiltonian $\mc H_+$ describes a single gapless phonon mode corresponding to fluctuations of the total charge $\partial_x\phi_+=\partial_x(\phi_0+\phi_2)$. %On the other hand 
$\mc H_-$ is generically gapped by the cosine terms; it realizes one of two distinct phases separated by a critical point. Which of the two phases is established depends on which one of the two cosine terms is larger and dominates the physics.

The {\em `trivial'} phase is established when the interaction dominates and $\t_-$ is pinned to $0$ or $\pi$ by the corresponding cosine term. This phase is adiabatically connected to the conventional spin-gapped Luther-Emery liquid, which forms in a one dimensional Fermi gas with spin symmetry and attractive interactions. In our case, the spin symmetry is broken by the Zeeman and spin-orbit couplings. But because the Zeeman field can only change the  spin by integer values, while the total spin can be either integer or half integer, it leaves a residual $Z_2$ fermion parity symmetry intact. The spin gap of the Luther-Emery liquid thus persists, in our case, as a gap to half-integer spin excitations, which carry the aforementioned $Z_2$ quantum number.

The phase we term {\em `topological'} is formed when the Zeeman field dominates over the interaction and pins the field $\phi_-$ to $0$ or to $\pi$, while the field  $\t_-$ is strongly fluctuating. %This state is adiabatically connected to simple non-interacting Fermions with the chemical potential inside the Zeeman gap. 
In contrast to the trivial phase, here single fermions, or half-integer spin excitations, are gapless.

One can drive a transition between the two phases by changing the value of the Zeeman coupling while keeping the chemical potential fixed (say at $\mu=0$). Alternatively, changing the chemical potential while keeping the Zeeman field fixed will have the same effect.  Specifically, tuning the chemical potential away from $\mu=0$ moves the putative inner Fermi points away from $q=0$, thereby making the Zeeman coupling less relevant. % and reducing its effective value. 
For weak attractive interactions the transition from the topological to the trivial phase is expected to occur approximately when the chemical potential goes above or below the Zeeman gap in the single particle dispersion.

\noindent {\em Zero modes.--- } We now turn to discuss an inhomogeneous system with spatial interfaces between the different phases discussed above. Such a situation occurs naturally in the harmonic trap potential as illustrated in Fig. \ref{fig:config}.a.
First, we discuss the ground state degeneracies expected to occur in such configurations based only on the properties of the low energy theory (\ref{H+}) and (\ref{H-}). Back-scattering terms, which are %necessarily generated 
present due to the absence of translational invariance, will be discussed later.

%Within the Bosonized representation, 
Within the trivial regions, the $\theta_-$ field is pinned to either $0$ or $\pi$. %each of the two gapped phases of ${\mc H}_-$ exhibits a fictitious double degeneracy of the ground state. For example, in the {\em 'trivial'} phase the field $\t_-$ can be pinned to $0$ or to $\pi$. The fact that this is a fictitious degeneracy becomes clear when
The two possible values of $\theta_-$ do not correspond to physically different states, however, since the value of $\t_-$ does not correspond to a physical observable. %gauge invariant (e.g., taking $R_0\rightarrow -R_0$ and $L_0 \rightarrow -L_0$ corresponds to $\t_- \rightarrow \t_- + \pi$). %Only the differences of $\t_-$ between distinct trivial regions are physical. 
In terms of the fermion densities, $\t_-$ can be written as
%we write this operator explicitly in terms of the physical fermion densities:
\bea
\t_-(x)&=&\pi \int_{-\infty}^x dx'(R_0 ^\dag R_0 - L_0 ^\dag L_0-R_2 ^\dag R_2+L_2 ^\dag L_2 ) \nn\\
&=& \pi\int_{-\infty}^x dx' [n_\ua(x)  - n_\da(x) ],
\eea
i.e. it is related to the total spin to the left of point $x$. Hence, only the differences between two points $\t_-(x_1)-\t_-(x_2)$ is a %gauge invariant 
physical quantity, independent of an arbitrary number of spins added at $-\infty$. Moreover, since the total spin is conserved only up to an integer value, the physical observable that can distinguish different ground states is the two point correlation function $P_{1,2} = \av{\cos(\t_-(x_2)-\t_-(x_1))}$. % = (-1)^{N_{1,2;\uparrow} - N_\downarrow}$, 
%which counts the %parity, or ``integerness'', of 
%the difference of the number of fermions with spin up and down 
%where $N_{\sigma = \uparrow,\downarrow}$ is the number of fermions with spin $\sigma$ located between $x_1$ and $x_2$. We will refer to $P_{1,2}$ as the \emph{spin parity} between $x_1$ and $x_2$. 
In a single region of the trivial phase the field $\t_-$ is pinned uniformly, and this correlation function is always unity.

If there are multiple trivial regions separated by topological regions (as in Fig. \ref{fig:config}), then
there can be multiple configurations of $\t_-(j)$  (defined in the trivial regions $j$), which represent physically distinct ground states. These states can be labeled by the values of the spin-parity on each of the topological regions: $P_{j,j+1}\equiv \av{\cos(\t_-(j+1)-\t_-(j)) }=\pm 1$. Note that a configuration with $P_{j,j+1}=-1$ requires the phase $\t_-$ to twist by $\pi$ inside the topological region between $x_j$ and $x_{j+1}$. Such a twist incurs an energy cost that is exponentially small in the size of the region, because the phase $\t_-$ has vanishing stiffness in the topological phase. Hence the different configurations of $P_{j,j+1}$ indeed represent zero modes up to the exponential splitting. Moreover the ground state degeneracy scales with the number of interfaces, $I$, as $2^{I/2}$, as expected from Majorana zero modes. In the supplementary material a direct mapping to Majorana fermions is derived by re-fermionizing the low energy theory.

It is important to realize, however, that the exact number of ground states is $2^{I/2}/2$, a factor of half smaller than for usual Majorana zero modes. The lost zero mode is the one that corresponds to the total parity of the system. States that differ in the total parity $P=\av{\cos[\pi(N_\ua -N_\da)]}$ must also differ in the total number $N=N_\ua+N_\da$. In a charge conserving system changing $N$ by one (equivalently, twisting the field $\phi_+$ by $\pi$ across the whole system) incurs an energy cost of $1/(\kappa L)$, where $\kappa^{-1} = u_+/K_+$ is the compressibility. Hence the putative zero mode is indistinguishable from usual phonons in this case.  There is no topological degeneracy in a configuration with a single topological region. This is an essential difference, also pointed out in Refs.  \cite{Fidkowski2011,Sau2011}, between a charge conserving system and one with proximity coupling to a superconductor.

{\em Back-scattering.---}
The spatial potential variations, which give rise to interfaces, may also lead to scattering between the four modes near the Fermi surface. Of the different processes, only scattering which involve  $2k_0$ momentum transfer, i.e.
$H_{bs} \sim e^{i 2k_0 x} V_{2k_0}R_2^\dag L_0 +  e^{-i 2k_0 x} V_{2k_0} L_2^\dag R_0 + \text{H.c.}$, can be potentially harmful to the topological degeneracy discussed above. 

These terms have a simple interpretation when expressed in terms of the bosonic fields. The wire can then be viewed as a thin superconducting strip, and the backscattering corresponds to tunneling of an $hc/2e$ vortex across it.  A second order process involving a vortex tunneling across the parts of the wire on the two sides of a topological segment is equivalent to a vortex encircling that region (see also Ref. \cite{Fidkowski2011}). Such a process cannot change the fermion parity of the topological segment, but it acquires a sign which depends on the fermion parity there. This leads to a splitting of the degeneracy proportional to the magnitude of the second order process. Because the process involves excitation of the symmetric (charge) sector, the splitting is proportional to the charge correlations between the two scattering points $\av{\exp[i(\phi_+(x_1)-\phi_+(x_2)]}$, i.e.  $\D E \sim u_+^{-1}V_{bs}(x_1)V_{bs}(x_2)|x_1-x_2|^{-K_+/4}$.
%Depending on the value of $K_+$ this term can give rise to two distinct behaviors:
%(i) For $K_+<4$ the scattering potential is a relevant perturbation and thus grows in the RG sense until it disconnects the two sides of the gas. If there is a distribution of such scatterers they will eventually separate all topological regions into isolated islands, which will give rise to a ground state energy splitting given by the charging energy associated by transferring particles between the islands. Therefore, in this case the ground state energy splitting will generally scale as $1/L$.
%(ii) For $K_+>4$ the scatterer becomes a weak-link through which vortices can tunnel across and cause $\pi$ phase slips in the superconducting phase $\t_+$. As shown by Ref.~\cite{Fidkowski2011} and in the SI, these phase slips give rise to a diagonal splitting in the ground state manifold which generally scales as $1/L^{K_+/2}$. The intutive picture is that two phase slip events occurring at the edges of a topological region are equivalent to the process in which a vortex goes around it. In this process the vortex acquires a Berry phase of $e^{\pm i\pi N}$ where $N$ is the number of Fermions in the region and thus performs a measurement of the parity in the region.

We are particularly interested in the backscattering produced by the smooth potential variation in a harmonic trap. We anticipate that in this case the backscattering matrix elements $V_{bs}$, induced by the potential gradient, would themselves  be functions of the system size, thus  leading to further reduction of the energy splitting.  To derive the magnitude of these %single particle 
scattering terms we consider the non-interacting part of the Hamiltonian (\ref{H_m}). First note that far from classical turning points the Fermi wavelength of both propagating modes is small compared to the %modulation 
rate of change of the underlying potential. In such a case the Wannier-Kramers-Brillouin (WKB) approximation may be applied. %and predicts that 
In this situation, any backscattering amplitude is exponentially small in the slope of the potential~\cite{Berry1972} (and therefore also in the total system size). The WKB approximation %is invalidated 
breaks down at the classical turning points, which %in this case 
are located where the chemical potential %just touches 
crosses the top or bottom of the Zeeman gap, or at the edges of the cloud where the chemical potential is at the bottom of the single particle dispersion. Consider a pair of backscattering events occurring on the two sides of a single topological region in the setup shown in Fig. \ref{fig:config}(a). %On one side of the region there is always strong backscattering from the edge of the cloud. 
On one side of this region there is always strong scattering from the point $x_0$ at the edge of the cloud.  %The other event must then occur 
The topological degeneracy is then lifted by backscattering events near the interfaces at $x_2$ or $x_3$.  

To compute the backscattering strength near %these interfaces, 
$x_{2,3}$ we linearize the potential near those turning points,  taking $V(x)\approx V(x_j)+r\,(x-x_j)$, where $r=m\Omega^2\,x_j\propto \Omega$. % is the slope.
Next, we apply a Gauge transformation, which transforms the linear potential into a vector potential linearly increasing in time, i.e. $V(x)\rightarrow V(x)-r\,x$ and $A(t) \rightarrow r\,t$. This leads to the translationally invariant Hamiltonian
\be
{\mc H}_0 = {\left (p  -  r\, t \right )^2 \over 2m} - \a \left(p  - r\, t \right) \s^x -\d_z \s^z, \label{H0}
\ee
where the momentum $p$ is conserved. %is now a good quantum number we can treat it as a constant and 
We are left with the task of finding the evolution of a two-level system with a time-dependent Hamiltonian. %an effective two level system. 
This is just the famous Landau-Zenner problem. %with a linear field sweep at the rate $\a r$. 
The back-scattering process involves a transition between the two dispersion branches separated by the Zeeman gap $\d_z$. The backscattering amplitude is therefore obtained directly from the well known Landau-Zenner formula 
$
V_{bs}\sim e^{-\pi\d_z^2/(\hbar\a r)}\sim
 e^{-\Omega_0/\Omega}, 
$
where $\Omega$ is the trap frequency and $\Omega_0= \d_z^2/\sqrt{m\a^2(\mu-\d_z)/2}\,$ for $ \d_z \gg \rho_0\,U$ (see SI). The total system length is inversely proportional to the trap frequency: %This can also be expressed in terms of the total system length 
$L\propto 1/\Omega$.  We conclude that in presence of a smooth confining potential, such as a harmonic trap, the splitting of the topological degeneracy is %remains 
exponentially small in the system size, as in the non-charge conserving case. %Moreover, 
Note that the entire low-energy spectrum is nearly doubly degenerate. 

\begin{figure}
\centering
\includegraphics[width=1\linewidth]{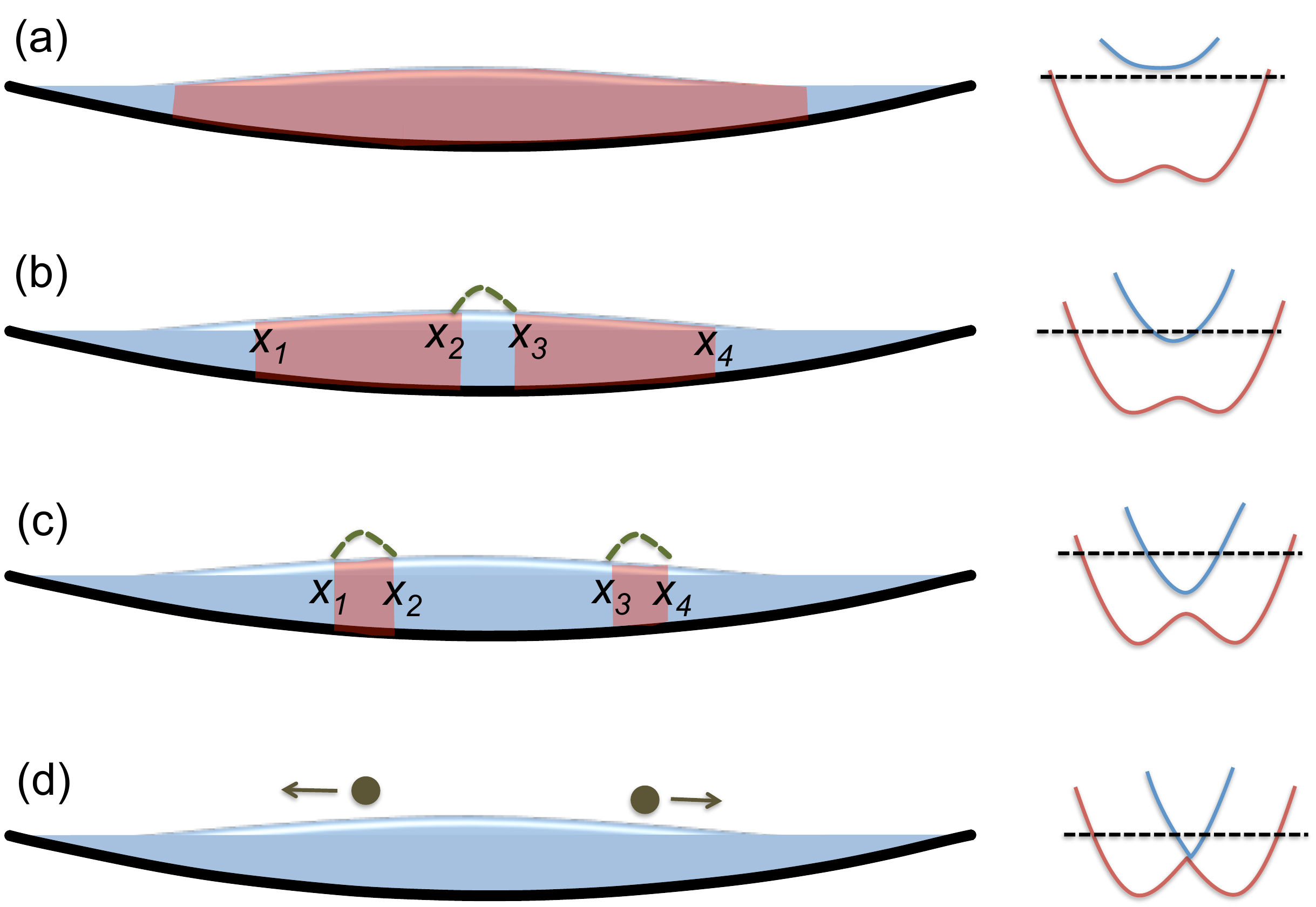}
\caption{{\em Probing scheme.} Change of trap configuration upon decreasing the Zeeman field from high value for the entire trap is in the topological phase (a) to zero field in which the entire trap is in the trivial phase (d). The topological pumping of quasiparticles which results from this process is described in the text.}\label{fig:pump}
\end{figure}

\noindent{\it Probing --} %Having identified degeneracies with topological origin in this setup we now turn to the question of detection. Earlier work, relying on a mean field (BCS) analysis, proposed to detect the degeneracy through emission spectra of single fermions by an RF pulse   \cite{Wei2012}. The hallmark of the topological state was to be the possibility to emit a fermion while leaving behind a zero energy hole. The problem with this approach is that a hole left behind in a charge conserving system will generically incur an energy which scales as $1/L$  and  thus its contribution to the spectral function would be indistinguishable from conventional elementary excitations of this gapless system.  
We propose to detect the topological state through a pumping effect induced by a slow variation of the Zeeman field, starting from a very high value down to zero. The configuration of segments in the trap slowly changes with the changing field as sketched in Fig. \ref{fig:pump}. (a) When $\d_z\gg \e_0$ the entire trap is in the {\em 'topological'} phase. As the field is reduced, trivial segments form at the two wings of the trap and expand inward. (b) A trivial segment is nucleated in the center and starts to expand while the two topological segments shrink. (c) The topological segments eventually vanish leaving the entire system in the trivial phase with a gap to single particle excitations. This process can lead to a rather surprising final state with clear topological origin. 

The topological degeneracy is first established after the trivial segment is nucleated in the middle of the trap at stage (b). This stage is described by creation of a pair of Majorana zero modes at the newly formed interfaces 2 and 3. Because they are created from the vacuum these modes are in a definite fusion channel (i.e. the trivial channel $i\g_2\g_3=1$). In this case the fermion parities $P_{12}=i\g_1\g_2$  and similarly $P_{23}$ in the topological region cannot be definite. Rather the system must be in the superposition state $\ket{\psi_e}=\ket{1,-1}+\ket{-1,1}$ if the total particle number is odd, or in $\ket{\psi_o}=\ket{-1,-1}+\ket{1,1}$ if it is even. In the subsequent evolution, while all interfaces are well separated, the system remains frozen in  this state (we assume the sweep is fast compared to the exponentially slow dynamics within the topological subspace). 

Finally the two pairs $\g_1,\g_2$ and $\g_3,\g_4$ fuse when the topological regions shrink and vanish. In the case of odd total particle number this process will always end up creating a single fermion quasiparticle (originating from the odd parity state of either the left or right topological region) above the parity gap of the trivial state. This is expected and does not rely on the topological properties of the intermediate phases. On the other hand, if the total number is even, then we have 50\% chance to end up with a pair of quasi-particles above the gap. Here there is a surprise in the apparent contradiction with naive application of the adiabatic which would suggest that we should end up in the ground state. 
When probing an ensemble of systems we expect half of them to have even particle number and half odd. Therefore the  average energy quasi-particles created in a sweep is $\D$ (where $\D$ is the parity gap) in contrast with the naive expectation of $\D/2$. We expect this scheme to work even at finite temperature as long as the final temperature $T_{\text{f}}\ll \D$ because the topological degeneracy persists to the low lying collective excitations.

\noindent{\it Conclusions --}
We predicted that an ultra-cold Fermi gas with Rashba-like spin orbit coupling in a one dimensional harmonic trap, a Zeeman field and intrinsic attractive interactions will form a novel topological state. Majorana zero modes are associated with interfaces between topological and trivial phases in the trap  which form in different regions due to spatial modulation of the potential. One important difference of the charge conserving system from one that is proximity coupled to a superconductor is that in the former case there is no degeneracy associated with a single topological segment; at least two are needed for a non-trivial degeneracy. 
We have proposed a simple pumping scheme that would detect the degeneracy and expose its topological origin.

\noindent{\em Acknowledgements --} We thank Benjamin Lev, 	
Sylvain Nascimbene, Xiao-Liang Qi, Jonathan Schattner and Yuval Oreg for helpful discussions. EA and EB thank the Aspen center for physics under NSF Grant \#1066293 for hospitality.
This research was supported in part by the ERC synergy grant UQUAM (EA and JR), the ISF (EA and EB) the Minerva Foundation (EA and EB). 

\bibliographystyle{apsrev} 

\begin{thebibliography}{18}%
\makeatletter
\providecommand \@ifxundefined [1]{%
 \@ifx{#1\undefined}
}%
\providecommand \@ifnum [1]{%
 \ifnum #1\expandafter \@firstoftwo
 \else \expandafter \@secondoftwo
 \fi
}%
\providecommand \@ifx [1]{%
 \ifx #1\expandafter \@firstoftwo
 \else \expandafter \@secondoftwo
 \fi
}%
\providecommand \natexlab [1]{#1}%
\providecommand \enquote  [1]{``#1''}%
\providecommand \bibnamefont  [1]{#1}%
\providecommand \bibfnamefont [1]{#1}%
\providecommand \citenamefont [1]{#1}%
\providecommand \href@noop [0]{\@secondoftwo}%
\providecommand \href [0]{\begingroup \@sanitize@url \@href}%
\providecommand \@href[1]{\@@startlink{#1}\@@href}%
\providecommand \@@href[1]{\endgroup#1\@@endlink}%
\providecommand \@sanitize@url [0]{\catcode `\\12\catcode `\$12\catcode
  `\&12\catcode `\#12\catcode `\^12\catcode `\_12\catcode `\%12\relax}%
\providecommand \@@startlink[1]{}%
\providecommand \@@endlink[0]{}%
\providecommand \url  [0]{\begingroup\@sanitize@url \@url }%
\providecommand \@url [1]{\endgroup\@href {#1}{\urlprefix }}%
\providecommand \urlprefix  [0]{URL }%
\providecommand \Eprint [0]{\href }%
\providecommand \doibase [0]{http://dx.doi.org/}%
\providecommand \selectlanguage [0]{\@gobble}%
\providecommand \bibinfo  [0]{\@secondoftwo}%
\providecommand \bibfield  [0]{\@secondoftwo}%
\providecommand \translation [1]{[#1]}%
\providecommand \BibitemOpen [0]{}%
\providecommand \bibitemStop [0]{}%
\providecommand \bibitemNoStop [0]{.\EOS\space}%
\providecommand \EOS [0]{\spacefactor3000\relax}%
\providecommand \BibitemShut  [1]{\csname bibitem#1\endcsname}%
\let\auto@bib@innerbib\@empty
%</preamble>
\bibitem [{\citenamefont {Mourik}\ \emph {et~al.}(2012)\citenamefont {Mourik},
  \citenamefont {Zuo}, \citenamefont {Frolov}, \citenamefont {Plissard},
  \citenamefont {Bakkers},\ and\ \citenamefont {Kouwenhoven}}]{Mourik2012}%
  \BibitemOpen
  \bibfield  {author} {\bibinfo {author} {\bibfnamefont {V.}~\bibnamefont
  {Mourik}}, \bibinfo {author} {\bibfnamefont {K.}~\bibnamefont {Zuo}},
  \bibinfo {author} {\bibfnamefont {S.~M.}\ \bibnamefont {Frolov}}, \bibinfo
  {author} {\bibfnamefont {S.~R.}\ \bibnamefont {Plissard}}, \bibinfo {author}
  {\bibfnamefont {E.~P. a.~M.}\ \bibnamefont {Bakkers}}, \ and\ \bibinfo
  {author} {\bibfnamefont {L.~P.}\ \bibnamefont {Kouwenhoven}},\ }\href
  {\doibase 10.1126/science.1222360} {\bibfield  {journal} {\bibinfo  {journal}
  {Science (New York, N.Y.)}\ }\textbf {\bibinfo {volume} {336}},\ \bibinfo
  {pages} {1003} (\bibinfo {year} {2012})}\BibitemShut {NoStop}%
\bibitem [{\citenamefont {Das}\ \emph {et~al.}(2012)\citenamefont {Das},
  \citenamefont {Ronen}, \citenamefont {Most}, \citenamefont {Oreg},
  \citenamefont {Heiblum},\ and\ \citenamefont {Shtrikman}}]{Das2012}%
  \BibitemOpen
  \bibfield  {author} {\bibinfo {author} {\bibfnamefont {A.}~\bibnamefont
  {Das}}, \bibinfo {author} {\bibfnamefont {Y.}~\bibnamefont {Ronen}}, \bibinfo
  {author} {\bibfnamefont {Y.}~\bibnamefont {Most}}, \bibinfo {author}
  {\bibfnamefont {Y.}~\bibnamefont {Oreg}}, \bibinfo {author} {\bibfnamefont
  {M.}~\bibnamefont {Heiblum}}, \ and\ \bibinfo {author} {\bibfnamefont
  {H.}~\bibnamefont {Shtrikman}},\ }\href {\doibase 10.1038/nphys2479}
  {\bibfield  {journal} {\bibinfo  {journal} {Nature Physics}\ }\textbf
  {\bibinfo {volume} {8}},\ \bibinfo {pages} {887} (\bibinfo {year}
  {2012})}\BibitemShut {NoStop}%
\bibitem [{\citenamefont {Oreg}\ \emph {et~al.}(2010)\citenamefont {Oreg},
  \citenamefont {Refael},\ and\ \citenamefont {von Oppen}}]{Oreg2010}%
  \BibitemOpen
  \bibfield  {author} {\bibinfo {author} {\bibfnamefont {Y.}~\bibnamefont
  {Oreg}}, \bibinfo {author} {\bibfnamefont {G.}~\bibnamefont {Refael}}, \ and\
  \bibinfo {author} {\bibfnamefont {F.}~\bibnamefont {von Oppen}},\ }\href
  {\doibase 10.1103/PhysRevLett.105.177002} {\bibfield  {journal} {\bibinfo
  {journal} {Physical Review Letters}\ }\textbf {\bibinfo {volume} {105}},\
  \bibinfo {pages} {177002} (\bibinfo {year} {2010})}\BibitemShut {NoStop}%
\bibitem [{\citenamefont {Lutchyn}\ \emph {et~al.}(2010)\citenamefont
  {Lutchyn}, \citenamefont {Sau},\ and\ \citenamefont {{Das
  Sarma}}}]{Lutchyn2010}%
  \BibitemOpen
  \bibfield  {author} {\bibinfo {author} {\bibfnamefont {R.~M.}\ \bibnamefont
  {Lutchyn}}, \bibinfo {author} {\bibfnamefont {J.~D.}\ \bibnamefont {Sau}}, \
  and\ \bibinfo {author} {\bibfnamefont {S.}~\bibnamefont {{Das Sarma}}},\
  }\href {\doibase 10.1103/PhysRevLett.105.077001} {\bibfield  {journal}
  {\bibinfo  {journal} {Physical Review Letters}\ }\textbf {\bibinfo {volume}
  {105}},\ \bibinfo {pages} {077001} (\bibinfo {year} {2010})}\BibitemShut
  {NoStop}%
\bibitem [{\citenamefont {Jiang}\ \emph {et~al.}(2011)\citenamefont {Jiang},
  \citenamefont {Kitagawa}, \citenamefont {Alicea}, \citenamefont {Akhmerov},
  \citenamefont {Pekker}, \citenamefont {Refael}, \citenamefont {Cirac},
  \citenamefont {Demler}, \citenamefont {Lukin},\ and\ \citenamefont
  {Zoller}}]{Jiang2011}%
  \BibitemOpen
  \bibfield  {author} {\bibinfo {author} {\bibfnamefont {L.}~\bibnamefont
  {Jiang}}, \bibinfo {author} {\bibfnamefont {T.}~\bibnamefont {Kitagawa}},
  \bibinfo {author} {\bibfnamefont {J.}~\bibnamefont {Alicea}}, \bibinfo
  {author} {\bibfnamefont {A.~R.}\ \bibnamefont {Akhmerov}}, \bibinfo {author}
  {\bibfnamefont {D.}~\bibnamefont {Pekker}}, \bibinfo {author} {\bibfnamefont
  {G.}~\bibnamefont {Refael}}, \bibinfo {author} {\bibfnamefont {J.~I.}\
  \bibnamefont {Cirac}}, \bibinfo {author} {\bibfnamefont {E.}~\bibnamefont
  {Demler}}, \bibinfo {author} {\bibfnamefont {M.~D.}\ \bibnamefont {Lukin}}, \
  and\ \bibinfo {author} {\bibfnamefont {P.}~\bibnamefont {Zoller}},\ }\href
  {\doibase 10.1103/PhysRevLett.106.220402} {\bibfield  {journal} {\bibinfo
  {journal} {Phys. Rev. Lett.}\ }\textbf {\bibinfo {volume} {106}},\ \bibinfo
  {pages} {220402} (\bibinfo {year} {2011})}\BibitemShut {NoStop}%
\bibitem [{\citenamefont {Lin}\ \emph {et~al.}(2011)\citenamefont {Lin},
  \citenamefont {Jim\'{e}nez-Garc\'{\i}a},\ and\ \citenamefont
  {Spielman}}]{Lin2011}%
  \BibitemOpen
  \bibfield  {author} {\bibinfo {author} {\bibfnamefont {Y.-J.}\ \bibnamefont
  {Lin}}, \bibinfo {author} {\bibfnamefont {K.}~\bibnamefont
  {Jim\'{e}nez-Garc\'{\i}a}}, \ and\ \bibinfo {author} {\bibfnamefont {I.~B.}\
  \bibnamefont {Spielman}},\ }\href {\doibase 10.1038/nature09887} {\bibfield
  {journal} {\bibinfo  {journal} {Nature}\ }\textbf {\bibinfo {volume} {471}},\
  \bibinfo {pages} {83} (\bibinfo {year} {2011})}\BibitemShut {NoStop}%
\bibitem [{\citenamefont {Cheuk}\ \emph {et~al.}(2012)\citenamefont {Cheuk},
  \citenamefont {Sommer}, \citenamefont {Hadzibabic}, \citenamefont {Yefsah},
  \citenamefont {Bakr},\ and\ \citenamefont {Zwierlein}}]{Cheuk2012}%
  \BibitemOpen
  \bibfield  {author} {\bibinfo {author} {\bibfnamefont {L.~W.}\ \bibnamefont
  {Cheuk}}, \bibinfo {author} {\bibfnamefont {A.~T.}\ \bibnamefont {Sommer}},
  \bibinfo {author} {\bibfnamefont {Z.}~\bibnamefont {Hadzibabic}}, \bibinfo
  {author} {\bibfnamefont {T.}~\bibnamefont {Yefsah}}, \bibinfo {author}
  {\bibfnamefont {W.~S.}\ \bibnamefont {Bakr}}, \ and\ \bibinfo {author}
  {\bibfnamefont {M.~W.}\ \bibnamefont {Zwierlein}},\ }\href {\doibase
  10.1103/PhysRevLett.109.095302} {\bibfield  {journal} {\bibinfo  {journal}
  {Physical Review Letters}\ }\textbf {\bibinfo {volume} {109}},\ \bibinfo
  {pages} {095302} (\bibinfo {year} {2012})}\BibitemShut {NoStop}%
\bibitem [{\citenamefont {Wang}\ \emph {et~al.}(2012)\citenamefont {Wang},
  \citenamefont {Yu}, \citenamefont {Fu}, \citenamefont {Miao}, \citenamefont
  {Huang}, \citenamefont {Chai}, \citenamefont {Zhai},\ and\ \citenamefont
  {Zhang}}]{Wang2012}%
  \BibitemOpen
  \bibfield  {author} {\bibinfo {author} {\bibfnamefont {P.}~\bibnamefont
  {Wang}}, \bibinfo {author} {\bibfnamefont {Z.-Q.}\ \bibnamefont {Yu}},
  \bibinfo {author} {\bibfnamefont {Z.}~\bibnamefont {Fu}}, \bibinfo {author}
  {\bibfnamefont {J.}~\bibnamefont {Miao}}, \bibinfo {author} {\bibfnamefont
  {L.}~\bibnamefont {Huang}}, \bibinfo {author} {\bibfnamefont
  {S.}~\bibnamefont {Chai}}, \bibinfo {author} {\bibfnamefont {H.}~\bibnamefont
  {Zhai}}, \ and\ \bibinfo {author} {\bibfnamefont {J.}~\bibnamefont {Zhang}},\
  }\href {\doibase 10.1103/PhysRevLett.109.095301} {\bibfield  {journal}
  {\bibinfo  {journal} {Phys. Rev. Lett.}\ }\textbf {\bibinfo {volume} {109}},\
  \bibinfo {pages} {095301} (\bibinfo {year} {2012})}\BibitemShut {NoStop}%
\bibitem [{\citenamefont {Cui}\ \emph {et~al.}(2013)\citenamefont {Cui},
  \citenamefont {Lian}, \citenamefont {Ho}, \citenamefont {Lev},\ and\
  \citenamefont {Zhai}}]{Cui2013}%
  \BibitemOpen
  \bibfield  {author} {\bibinfo {author} {\bibfnamefont {X.}~\bibnamefont
  {Cui}}, \bibinfo {author} {\bibfnamefont {B.}~\bibnamefont {Lian}}, \bibinfo
  {author} {\bibfnamefont {T.-L.}\ \bibnamefont {Ho}}, \bibinfo {author}
  {\bibfnamefont {B.~L.}\ \bibnamefont {Lev}}, \ and\ \bibinfo {author}
  {\bibfnamefont {H.}~\bibnamefont {Zhai}},\ }\href {\doibase
  10.1103/PhysRevA.88.011601} {\bibfield  {journal} {\bibinfo  {journal} {Phys.
  Rev. A}\ }\textbf {\bibinfo {volume} {88}},\ \bibinfo {pages} {011601}
  (\bibinfo {year} {2013})}\BibitemShut {NoStop}%
\bibitem [{\citenamefont {Fidkowski}\ \emph {et~al.}(2011)\citenamefont
  {Fidkowski}, \citenamefont {Lutchyn}, \citenamefont {Nayak},\ and\
  \citenamefont {Fisher}}]{Fidkowski2011}%
  \BibitemOpen
  \bibfield  {author} {\bibinfo {author} {\bibfnamefont {L.}~\bibnamefont
  {Fidkowski}}, \bibinfo {author} {\bibfnamefont {R.~M.}\ \bibnamefont
  {Lutchyn}}, \bibinfo {author} {\bibfnamefont {C.}~\bibnamefont {Nayak}}, \
  and\ \bibinfo {author} {\bibfnamefont {M.~P.~A.}\ \bibnamefont {Fisher}},\
  }\href {\doibase 10.1103/PhysRevB.84.195436} {\bibfield  {journal} {\bibinfo
  {journal} {Physical Review B}\ }\textbf {\bibinfo {volume} {84}},\ \bibinfo
  {pages} {195436} (\bibinfo {year} {2011})}\BibitemShut {NoStop}%
\bibitem [{\citenamefont {Sau}\ \emph {et~al.}(2011)\citenamefont {Sau},
  \citenamefont {Halperin}, \citenamefont {Flensberg},\ and\ \citenamefont
  {{Das Sarma}}}]{Sau2011}%
  \BibitemOpen
  \bibfield  {author} {\bibinfo {author} {\bibfnamefont {J.~D.}\ \bibnamefont
  {Sau}}, \bibinfo {author} {\bibfnamefont {B.~I.}\ \bibnamefont {Halperin}},
  \bibinfo {author} {\bibfnamefont {K.}~\bibnamefont {Flensberg}}, \ and\
  \bibinfo {author} {\bibfnamefont {S.}~\bibnamefont {{Das Sarma}}},\ }\href
  {\doibase 10.1103/PhysRevB.84.144509} {\bibfield  {journal} {\bibinfo
  {journal} {Physical Review B}\ }\textbf {\bibinfo {volume} {84}},\ \bibinfo
  {pages} {144509} (\bibinfo {year} {2011})}\BibitemShut {NoStop}%
\bibitem [{\citenamefont {Wei}\ and\ \citenamefont {Mueller}(2012)}]{Wei2012}%
  \BibitemOpen
  \bibfield  {author} {\bibinfo {author} {\bibfnamefont {R.}~\bibnamefont
  {Wei}}\ and\ \bibinfo {author} {\bibfnamefont {E.~J.}\ \bibnamefont
  {Mueller}},\ }\href {\doibase 10.1103/PhysRevA.86.063604} {\bibfield
  {journal} {\bibinfo  {journal} {Physical Review A}\ }\textbf {\bibinfo
  {volume} {86}},\ \bibinfo {pages} {063604} (\bibinfo {year}
  {2012})}\BibitemShut {NoStop}%
\bibitem [{\citenamefont {Liu}\ and\ \citenamefont {Hu}(2012)}]{Liu2012}%
  \BibitemOpen
  \bibfield  {author} {\bibinfo {author} {\bibfnamefont {X.-J.}\ \bibnamefont
  {Liu}}\ and\ \bibinfo {author} {\bibfnamefont {H.}~\bibnamefont {Hu}},\
  }\href {\doibase 10.1103/PhysRevA.85.033622} {\bibfield  {journal} {\bibinfo
  {journal} {Physical Review A}\ }\textbf {\bibinfo {volume} {85}},\ \bibinfo
  {pages} {033622} (\bibinfo {year} {2012})}\BibitemShut {NoStop}%
\bibitem [{Note1()}]{Note1}%
  \BibitemOpen
  \bibinfo {note} {Linear coupling terms $\partial _x {\theta }_- \partial _x
  {\theta }_+$ and $\partial _x \phi _- \partial _x \phi _+$ generated by the
  canonical transformation are irrelevant in the long wavelength limit because
  the mode (${\theta }_-,\phi _-$) gapped (except at a critical
  point).}\BibitemShut {Stop}%
\bibitem [{\citenamefont {Berry}\ and\ \citenamefont
  {Mount}(1972)}]{Berry1972}%
  \BibitemOpen
  \bibfield  {author} {\bibinfo {author} {\bibfnamefont {M.~V.}\ \bibnamefont
  {Berry}}\ and\ \bibinfo {author} {\bibfnamefont {K.~E.}\ \bibnamefont
  {Mount}},\ }\href@noop {} {\bibfield  {journal} {\bibinfo  {journal} {Reps.
  Prog. Phys}\ }\textbf {\bibinfo {volume} {35}},\ \bibinfo {pages} {315}
  (\bibinfo {year} {1972})}\BibitemShut {NoStop}%
\bibitem [{\citenamefont {Kane}\ and\ \citenamefont {Fisher}(1992)}]{Kane1992}%
  \BibitemOpen
  \bibfield  {author} {\bibinfo {author} {\bibfnamefont {C.~L.}\ \bibnamefont
  {Kane}}\ and\ \bibinfo {author} {\bibfnamefont {M.~P.~A.}\ \bibnamefont
  {Fisher}},\ }\href {\doibase 10.1103/PhysRevB.46.15233} {\bibfield  {journal}
  {\bibinfo  {journal} {Phys. Rev. B}\ }\textbf {\bibinfo {volume} {46}},\
  \bibinfo {pages} {15233} (\bibinfo {year} {1992})}\BibitemShut {NoStop}%
\bibitem [{\citenamefont {Lecheminant}(2002)}]{Lecheminant2002}%
  \BibitemOpen
  \bibfield  {author} {\bibinfo {author} {\bibfnamefont {P.}~\bibnamefont
  {Lecheminant}},\ }\href {\doibase 10.1016/S0550-3213(02)00474-1} {\bibfield
  {journal} {\bibinfo  {journal} {Nuclear Physics B}\ }\textbf {\bibinfo
  {volume} {639}},\ \bibinfo {pages} {502} (\bibinfo {year}
  {2002})}\BibitemShut {NoStop}%
\bibitem [{\citenamefont {Sitte}\ \emph {et~al.}(2009)\citenamefont {Sitte},
  \citenamefont {Rosch}, \citenamefont {Meyer}, \citenamefont {Matveev},\ and\
  \citenamefont {Garst}}]{sitte2009a}%
  \BibitemOpen
  \bibfield  {author} {\bibinfo {author} {\bibfnamefont {M.}~\bibnamefont
  {Sitte}}, \bibinfo {author} {\bibfnamefont {A.}~\bibnamefont {Rosch}},
  \bibinfo {author} {\bibfnamefont {J.~S.}\ \bibnamefont {Meyer}}, \bibinfo
  {author} {\bibfnamefont {K.~A.}\ \bibnamefont {Matveev}}, \ and\ \bibinfo
  {author} {\bibfnamefont {M.}~\bibnamefont {Garst}},\ }\href {\doibase
  10.1103/PhysRevLett.102.176404} {\bibfield  {journal} {\bibinfo  {journal}
  {Physical Review Letters}\ }\textbf {\bibinfo {volume} {102}},\ \bibinfo
  {pages} {176404} (\bibinfo {year} {2009})}\BibitemShut {NoStop}%
\end{thebibliography}
\newpage

\appendix
\onecolumngrid
\section{\Large{\bf Supplementary Information}}

\section{Fermionization of the double sine-Gordon model and Majorana edge modes in the parabolic trap}
%In this section we obtain the zero modes wave-functions localized around the four interfaces illustrated in Fig.1.a of the main text. We first derive the critical theory separating the trivial and topological phases. We then use this theory to solve the Bogoliubov-deGennes equations for the case of the parabolic trap potential illustrated in Fig.1.a of the main text.

The low energy effective theory which describes the two possible phases of the superfluid in the trap and the critical point between them
can be written in terms of Majorana fermions through reformionization of the double sine-Gordon model (Eq. (3) in the main text) in a rather standard way (see e.g. \cite{Lecheminant2002}). We define the Majorana fermions:
\[\xi_R \sim \sqrt{\rho_0 \over \pi}\,\cos{(\t_- - \phi_-)} \;\;\;\;;\;\;\;\;\xi_L \sim \sqrt{\rho_0 \over \pi}\,\sin{(\t_- + \phi_-)}\]
\[\eta_R \sim \sqrt{\rho_0 \over \pi}\,\sin{(\t_- - \phi_-)} \;\;\;\;;\;\;\;\;\eta_L \sim \sqrt{\rho_0 \over \pi}\,\cos {(\t_- + \phi_-)}.\]
The two Majorana species $\xi$ and $\eta$ acquire the quadratic mass terms $-i\D_+\eta_L\eta_R-i\D_-\xi_L\xi_R$, where  $\D_\pm \equiv \D \pm \d_z$ and $\D\equiv \rho_0 U/\pi$. Hence, by varying the effective Zeeman field $\d_z$ we tune the pair of lower energy Majorana modes $\xi^T=(\xi_L,\xi_R)$ through a critical point at which their gap changes sign. The second pair of Majorana modes retains  a higher gap throughout, which makes the quartic interaction between the four Majorana species highly irrelevant and  allows to safely integrate out the fields $\eta_{L,R}$.
This leads to a low energy Hamiltonian of the form:
\begin{align}
&\mc H_+={u \over 2\pi}\int dx\left[ K_+(\partial_x \t_+)^2 + {1\over K_+}(\partial_x \phi_+)^2\right]\label{H+2} \\
&\mc H_-={1\over 2}\int dx \,\xi ^{\mrm T}\,\left[  v (-i\partial_x) \tau^z+\D_- \tau^y  \right]\,\xi \label{H-2}\\
&\mc H_{+ -}=i \lambda \int dx\, \partial_x \phi \, \xi_R \xi_L \label{H+-2}
\end{align}
This theory describes a transition between two different pairing phases in the fermion sector as the gap $\D_-$ changes sign.
The coupling term $\lambda$ can affect the critical point in an interesting way \cite{sitte2009a}, but it cannot change the two phases where the Majorana modes are gapped.
We note that this coupling is not obtained directly from the refermionization applied above, but is rather generated at higher order upon integrating out the higher energy modes. Hence the coupling constant $\lambda$ is aproiori small.

We now consider the parabolic trap potential $V(x)=\half m \Omega^2 x^2$, where $m$ is the bare mass and $\mu>\d_z$ so that the chemical potential in the center of the trap is located above Zeeman gap.
Within a local density approximation, we can think of the external potential as giving rise to a local chemical potential $\tilde{\mu}(x)=\mu-V(x)$, where of course the true electro-chemical potential is constant along the trap.
Decreasing the local chemical potential can be viewed instead as a modulation of the effective Zeeman field  $\d_z(x)$ in a constant chemical potential, as it becomes gradually less effective in polarizing the spin.
The effective value of the Zeeman field is determined by the expectation value of the Zeeman energy of particles at the fermi energy. Using the single particle dispersion and assuming that throughout the region of interest $V(x)\ll \epsilon_F\approx m \a^2 /2$, the effective Zeeman field is given by
\be
\tilde{\d}_z(x)\equiv\d_z\av{\s^z}\approx\d_z\left[ 1-\left({V(x)-\mu \over \d_Z}\right)^2 \right]
\ee
Note that in this expression we have assumed that $\D$ is sufficiently large such that $V(x)$ at the position of the bound state is small compared to the full Zeeman gap $\d_z$, the accurate expressions are found in Ref.[\onlinecite{Wei2012}].

The gap in the Majorana sector of the low energy theory [Eq. (\ref{H-2})] is determined by the difference between the pairing interaction and the effective Zeeman field $\D_-=\D-\tilde{\d}_z(x)$. The points in the trap where this function changes sign can now be given in terms of the bare Zeeman field and pairing interaction (The number conserving interaction $\Delta$ appears as a pairing field in the anti-symmetric, i.e. relative mode, sector):
\begin{align}
&x_{1,2}  =- \sqrt{2 \d_z \over m} \left[{\mu \over \d_z}\pm \sqrt{1-{\D \over \d_z}} \right]^{1/2} {1\over \Omega}\label{x12}\\
&x_{3,4}  = \sqrt{2 \d_z \over m} \left[{\mu \over \d_z}\mp \sqrt{1-{\D \over \d_z}} \right]^{1/2} {1\over \Omega}\,.\label{x34}
\end{align}
This result is consistent with that derived using a mean field (Bogoliubov-deGennes) treatment of the Fermi gas in a trap in Ref.[\onlinecite{Wei2012}].

To obtain the explicit form of the zero mode wave function we linearize $\D_-(x)$ around the four points $x_1\ldots x_4$:
$
\D_- (x)  \approx \pm{v}(x-x_j)/ d_j^2\,,
$
with
\[d_j^2  \equiv {v  \over   2m \sqrt {1-{\D\over\d_z}}  }{1\over \Omega^2 |x_j|}\,.\]
Using this gap function we get solution of the BdG equations near to the points where the gap changes sign is given by
\be
\chi_j (x) \sim \exp \left( {1\over v}\int_{0} ^{x} dx' \D_-(x'-x_j) \right) \approx  \left({1\over \pi d_j}\right)^{1/4}  e^{-\left({ x - x_j \over 2 d_j }\right)^2 }\label{chi}.
\ee
We see that the Majorana interface modes are localized on the length $d\propto \sqrt{L}\propto 1/\sqrt{\Omega}$. Hence the two interface modes become increasingly better defined as we take the limit of a larger trap  $\Omega\to 0$.
The bound state operator $\gamma_i$ and the gapped modes orthogonal to it $\bar{\gamma}_i$ are given by the transformation
\be
\begin{pmatrix} \g_j \\ \bar \g_j \end{pmatrix} = \int dx \,\chi_j(x)\begin{pmatrix} 1 & \eta_j \\ -\eta_j & 1\end{pmatrix}\begin{pmatrix} \xi_R(x) \\ \xi_L(x) \end{pmatrix}\label{bs}
\ee
where $\eta_j=\pm 1$ for the states at the opposite ends $j=1,2$ and $j=3,4$.

\section{Ground state energy splitting due to $2k_0$-backscattering}
In this section we discuss the ground-state energy splitting due to backscattering with $2k_0$ momentum transfer from local impurities in the potential. Such backscattering is described by the Hamiltonian
\[\mc H_{2k_0}(x)=V_{2k_0} e^{i2k_0 x}R^\dag_2 L_0 + \tilde V_{2k_0} e^{i2k_0 x} R^\dag_0 L_2 + \mrm{H.c.} \]
In terms of the bosonic fields
\bea
R_0(x)&\approx&F_0\sqrt{\rho_0 \over 2\pi }\,e^{i[\t_-(x)+\t_+(x) - \phi_-(x)] }\;\;\;\; ; \;\;\;\; L_0(x)\approx F_0\sqrt{\rho_0 \over 2\pi }\,e^{i[\t_-(x)+\t_+(x) + \phi_-(x)]}\nn\\
R_2(x)&\approx&F_2\sqrt{\rho_0 \over 2\pi }\,e^{i[\t_+(x) - \phi_+(x)+\phi_-(x)] }\;\;\;\; ; \;\;\;\; L_2(x)\approx F_2\sqrt{\rho_0 \over 2\pi }\,e^{i[\t_+(x) + \phi_+(x)-\phi_-(x)] }\nn
\eea
this Hamiltonian assumes the form
\be
\mc H_{2k_0}(x) = i F_2 F_0 g_{2k_0} \cos(\t_-(x)+\f) \cos(\phi_+(x)+2k_0 x) \label{H2k0}
\ee
where $\f$ is the phase between $V_{2k_0}$ and $\tilde V_{2k_0}$ and $F_0$, $F_2$ are sermonic Klein factors. The backscattering perturbation (\ref{H2k0}) thus couples the spin-parity operator $\cos\t_-(x)$ to an operator which generates $\pi$ phase slips in the phase $\t_+$. Therefore, this perturbation is irrelevant in the topological regions where the parity operator is disordered. However, as explained in the main text, in the trivial regions it may have a more harmful effect.

Depending on the value of $K_+$ it can give rise to two distinct behaviors:
(i) For $K_+<4$ the perturbation.~(\ref{H2k0}) is a relevant and thus effectively disconnects the two sides of the gas at the longest scales~\cite{Kane1992}. If there is a distribution of such scatterers they will eventually separate all topological regions into isolated islands, which will give rise to a ground state energy splitting given by the charging energy associated by transferring particles between the islands. In this case the ground state energy splitting will generally scale as $1/L$.
(ii) For $K_+>4$ the scatterer becomes a weak-link through which vortices can tunnel across and cause $\pi$ phase slips in the superconducting phase $\t_+$. The process in which a vortex encircles a given topological region is sensitive to the spin-parity in the region and thus splits the ground state degeneracy without modifying the spin-parity. To encircle the topological region the vortex must penetrate through the wire at two points located in the two trivial regions surrounding the topological one, and therefore this effect appears only at second order when there at least two scattering centers.

To see this we consider two scattering centers of the form (\ref{H2k0}) located at two points $x_1$ and $x_2$ surrounding a given topological region
\[\mc H_{bs} = \mc H_{2k_0}(x_1)+\mc H_{2k_0} (x_2) \]
By second order perturbation theory these terms give rise to a splitting
\be
\d E_{(2)}(P_{12}) = {|\langle H_{bs}\rangle|^2 \over \Lambda}=A+B\langle \cos \t_- (x_1)\cos \t_- (x_2)\rangle\langle \cos \phi_+ (x_1)\cos \phi_+ (x_2)\rangle = A+B\,P_{12}\, \left({1\over \rho_0|x_2-x_1|}\right)^{K_+\over 2}
\ee
where $A=|g_{2k_0}(x_1)|^2+|g_{2k_0}(x_2)|^2$, $B = 2\mrm{Re}[g_{2k_0}(x_1)g_{2k_0}(x_2)]\cos 2k_0 x \cos\f$ and $P_{12}$ is the parity in the topological region between $x_1$ and $x_2$. Therefore the splitting between the two parity states is given by
\be
\D E  = \d E_{(2)}( 1)-\d E_{(2)}( -1) = 2 B \left({1\over \rho_0|x_2-x_1|}\right)^{K_+\over 2} \label{dE}
\ee
Thus, this splitting is parametrically separated from the phonon excited states which scale as $1/L$. It is also important to point out that if one of the points $x_1$ or $x_2$ is located at the boundary with the vacuum where $\phi_+$ is pinned the exponent will be modified to $K_+/4$.

\section{$2k_0$-backscattering in the parabolic trap potential}

In this section we consider the $2k_0$-backscattering induced by the harmonic trap in the absence of any other impurity potential. We first assess the strength of the backscattering analytically by mapping the Hamiltonian of a free particle propagating in the harmonic potential to a Landau-Zenner sweep through the Zeeman gap. We then supplement this estimate with a numerical calculation.

As mentioned in the main text, the only points where the backscattering is potentially large enough to make a notable effect are the classical turning points where the group velocity goes to zero. In the case of Rashba dispersion there are two types of classical turning points. The first type occur at the very ends of the trap, at the interface with the vacuum. Since this type leads to backscattering only at the very ends of the wire it can not give rise to an energy splitting by Eq. (\ref{dE}).
The other type of turning points occur when the velocity of the inner modes $R_0,L_0$ goes to zero while the outer modes $R_2,L_2$ stay deep in the adiabatic limit. These are nothing but the interfaces between the topological and trivial regions in the trap.

\begin{figure*}
\centering
\includegraphics[width=11.5cm,height=5cm]{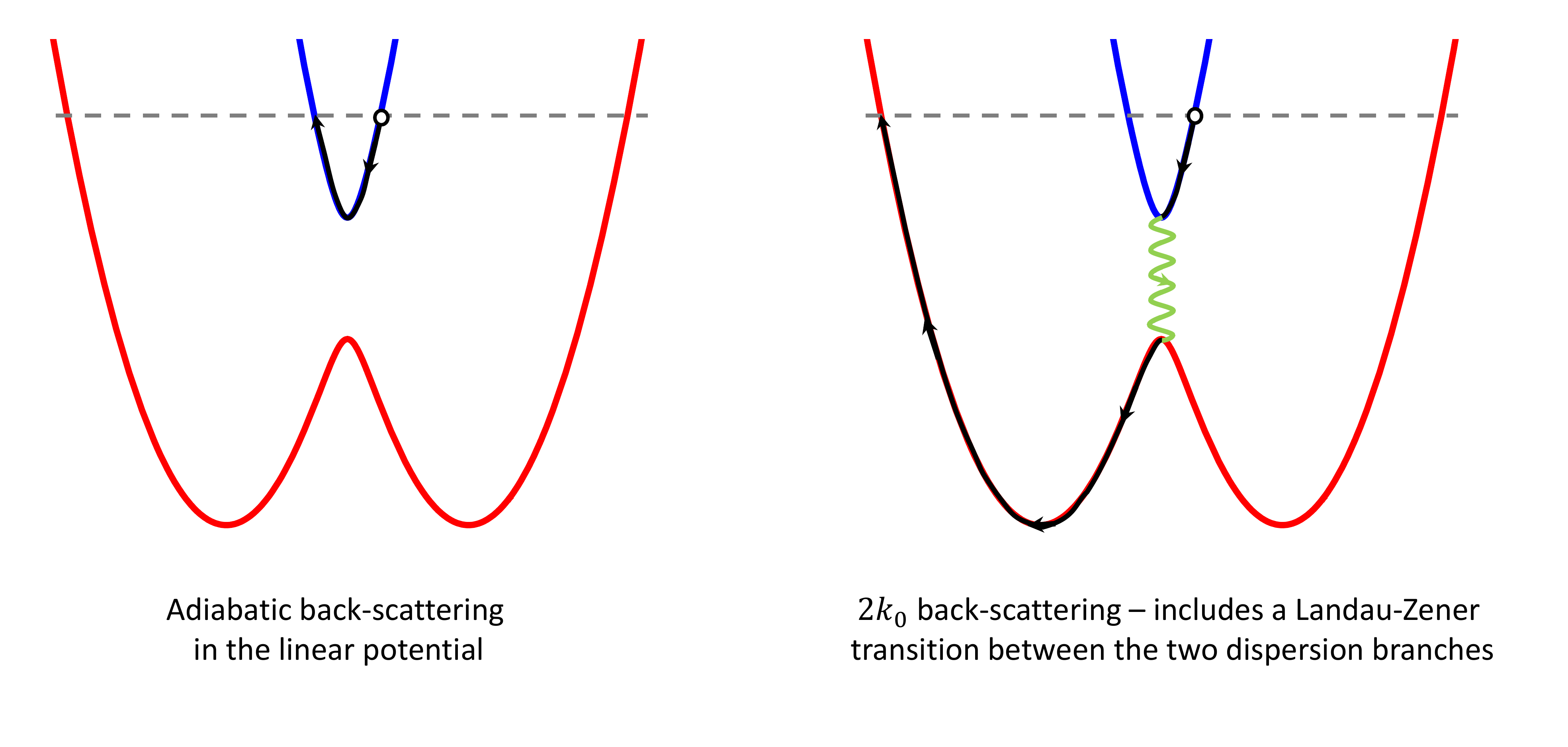}
\caption{The linear potential accelerates the momentum states linearly in time. Starting in an initial state in the inner branch the particle can either be backscattered adiabatically to negative momenta on the same branch (right) or undergo a Landau-Zener transition through the avoided crossing at $k=0$ set by the Zeeman gap.   }\label{fig:appendix1}
\end{figure*}

To study the backscattering near such a turning point located at $x=x_j$ we linearize the potential $V(x)\approx V(x_j)+r\,(x-x_j)+\ldots$, which leads to the Hamiltonian
\[\mc H_0 = {p^2 \over 2m}-\a p \s^x -\d_z \s^z+r\, (x-x_j) \,.\]
where $r = m \Omega^2 x_j$.
It is in convenient to use a gauge transformation
\[V\rightarrow  \tilde V= V-{\partial_t} f = 0  \;\;\; ;\;\;\; A \rightarrow \tilde A= \partial_x f = r\,t\]
where $f = r \,(x-x_j) \,t$ and $A$ is the vector potential. This leads to a translationally invariant Hamiltonian
\be
\tilde {\mc H}_0 = {\left (p  - rt \right )^2 \over 2m} - \a \left(p  - rt \right) \s^x -\d_z \s^z \, \label{H0}
\ee
In this representation it becomes clear that the spin-orbit coupling transforms the acceleration of the momentum states due to the linear potential into a linear sweep through an avoided crossing of magnitude $\d_z$ with a rate of $\a r$. In this case there are two distinct backscattering processes that may occur. Either the state of the particle follows adiabatically and backscatters within the same dispersion branch (see the left panel of Fig.~\ref{fig:appendix1}) or it undergoes a Landau-Zenner transition between the branches (right panel of Fig.~\ref{fig:appendix1}). The latter process is exactly the $2k_0$-backscattering and therefore it is bound from above by the Landau-Zenner probability
\be
\mc P_{LZ} \propto \exp\left[{-{2\pi\d_z^2 \over \a r}}\right]
\ee
where $r \propto \Omega \propto L^-1$. 

Let us consider for example the backscattering at the inner turning points $x_2$ and $x_3$ illustrated in Fig.1.a of the main text. In the limit of $\d_z \gg \rho_0 U$ the sweep rate is given by $\a r \approx  \a \Omega \sqrt{2 m (\mu - \d_z)}=2\Omega\sqrt{\e_0(\mu-\d_z) }$, where $\e_0 \equiv  m \a^2 /2$ (see Eq.~\ref{x12} and Eq.~\ref{x34}). Therefore overall the $2k_0$-backscattering probability is estimated by
\[ \mc P_{LZ} \sim \exp\left[-\pi \sqrt{ \d_z^2 \over \e_0 (\mu - \d_z) } {\d_z \over \Omega} \right] .\]
 
We now turn to estimate the backscattering by calculating the scattering matrix due to the barrier potential
\[ V(x) = {V_0\over 2}\left[ \tanh { {x\over 2a}}+1\right]\,.\]
In this potential the inner modes incur a classical turning point at $x=0$ where the slope of the potential is $V'(x_j) = {V_0\over 4 a}$. We seek a solution of the Schrodinger equation in terms of the scattering state 
\be
\Psi(x) = \bigg \{ \begin{matrix} & u(k_{F,2})\,e^{i\,k_{F,2}\,x}+r_{22}\,u(-k_{F,2})\,e^{-i\,k_{F,2}\,x}+r_{20}\,v(-k_{F,0})\,e^{-i\,k_{F,0}\,x}&\;,\;& x\rightarrow-\infty  \label{scattering state}\\ 
& t_{22}\,u(\tilde k_{F,2})\,e^{i\,\tilde k_{F,2}\,x}\;\;\;\,\;\;\;\;\;\;\;\;\;\;\;\;\;\;\;\;\;\;\;\;\;\;\;\;\;\;\;\;\;\;\;\;\;\;\;\;\;\;\;\;\;\;\;\;\;\;\;\;\;\;\;\;\;\;\;\;\;\;\;\;\;\;\;\;\;\;\;&\;,\;& x\rightarrow\infty \end{matrix}
\ee
where $k_{F,a}(\mu) = k_0\left[ 2+{\mu\over \e_0}\pm \sqrt{4+4{\mu \over \e_0} + {\d_z^2 \over \e_0^2}}\right]^{1/2}$ are the Fermi momenta at $x\rightarrow -\infty$, $\tilde k_{F,2} \equiv k_{F,a}(\mu-V_0)$ is the Fermi momentum of mode $a=2$ at $x\rightarrow \infty$, $\e_0 \equiv m\a^2/2$ and $k_0 = m\a$.  
$u(k)$ and $v(k)$ are the spinors of the lower and upper helical dispersion branches, respectively.

\begin{figure*}
\centering
\includegraphics[width=6cm,height=6.0cm]{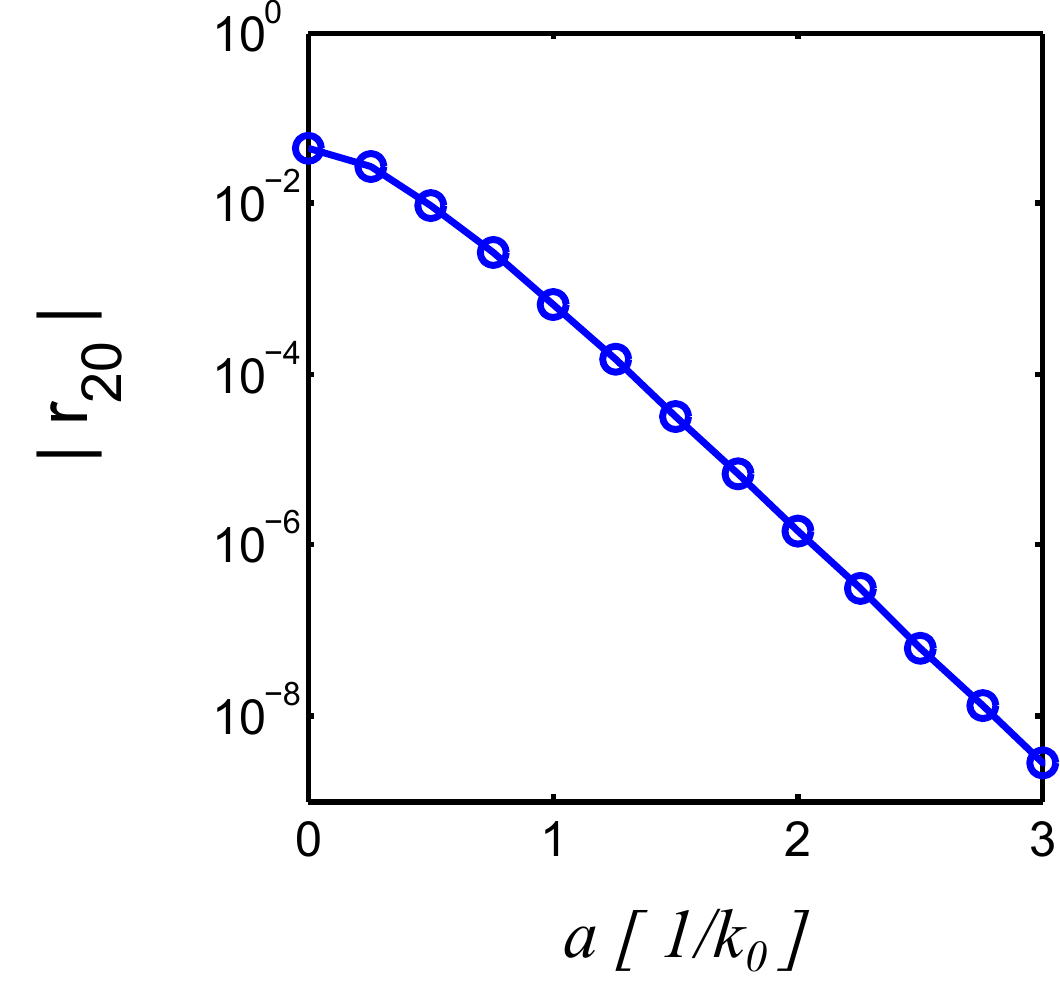}
\caption{The interband reflection coefficient $|r_{20}|$ as a function of the slope parameter $a$ for $\d_z = 0.25\,\e_0$, $V_0 = 0.2\,\e_0$, $\mu = 0.35\,\e_0$ and where we set $\e_0 = m \a^2 /2 =1$ and $k_0 = m \a = 1$. }\label{fig:appendix2}
\end{figure*}

The wave-function (\ref{scattering state}) describes a scattering state with a particle belonging to mode $a=2$ approaching the barrier from the right and scattering into all possible out-going states. The $2k_0$-backscattering amplitude is given by the reflection coefficient to band $a=0$, i.e. $r_{20}$.
In Fig.~\ref{fig:appendix2} we plot the magnitude of the interband reflection coefficient $|r_{20}|$ as a function of $a$ for  $\d_z = 0.25\,\e_0$, $V_0 = 0.2\,\e_0$ $\mu = 0.35\,\e_0$ and where we set $\e_0 = m \a^2 /2 =1$ and $k_0 = m \a = 1$. The figure shows clear exponential decay in the inverse slope in agreement with the Landau-Zenner argument.

\end{document}